%% file: arxiv.tex
\let\ORIGkernellabel\label   
\documentclass[
reprint,
showkeys,
superscriptaddress,
amsmath,amssymb,
aps,
pra,
]{revtex4-2}

\makeatletter
\AddToHook{package/nameref/before}{%
  \let\label\ORIGkernellabel
  \let\ltx@label\ORIGkernellabel}
\def\caption@documentclass{standard}
\makeatother

\usepackage[version=3]{mhchem} 

\usepackage{amsmath}
\usepackage{amssymb}
\usepackage{graphicx}
\usepackage{flushend} 
\usepackage{xcolor}
\usepackage{nicefrac}
\usepackage{dcolumn}
\usepackage{mathtools}

\DeclarePairedDelimiterX\braket[2]{\langle}{\rangle}{#1 \delimsize\vert #2}

\usepackage{siunitx}
\usepackage{amsfonts}
\usepackage{bm}
\usepackage[T1]{fontenc}

\usepackage{caption}
\usepackage{subcaption}

\usepackage{verbatim}
\usepackage{placeins}
\usepackage{booktabs}

\usepackage{orcidlink}

\newenvironment{strip}{\par\medskip}{\par\medskip}

\usepackage{hyperref}
\hypersetup{colorlinks=true, allcolors=black}

\newcommand{\JKUsemicon}{Semiconductor Physics Division, Institute of Semiconductor and Solid State Physics, Johannes Kepler University Linz, Altenberger Straße 69, 4040 Linz, Austria}

\newcommand{\IIT}{Department of Electrical Engineering, Indian Institute of Technology Delhi, 70174 New Delhi, India}

\newcommand{\UNICAMP}{Instituto de Física Gleb Wataghin, Universidade Estadual de Campinas (UNICAMP), 13083-859 Campinas, Brazil}

\newcommand{\JKUlightmatter}{Light-Matter Interaction Division, Institute of Semiconductor and Solid State Physics, Johannes Kepler University Linz, Altenberger Straße 69, 4040 Linz, Austria}

\newcommand{\UniS}{Institut für Halbleiteroptik und Funktionelle Grenzflächen, Center for Integrated Quantum Science and Technology (IQST) and SCoPE, University of Stuttgart, Allmandring 3, 70569 Stuttgart, Germany}

\begin{document}

\title{Planar metal-semiconductor Yagi-Uda type antennas for tunable narrow-linewidth quantum dot emitters}

\author{Ievgen Brytavskyi\,\orcidlink{0000-0003-4177-9818}}
\email{ievgen.brytavskyi@jku.at}
\affiliation{\JKUsemicon}

\author{Andreas Kitzmüller\,\orcidlink{0009-0007-2502-434X}}
\affiliation{\JKUsemicon}

\author{Santanu Manna\,\orcidlink{0000-0002-0946-2629}}
\affiliation{\IIT}

\author{Gabriel Undeutsch\,\orcidlink{0000-0001-5080-7695}}
\affiliation{\JKUsemicon}

\author{Christian Weidinger\,\orcidlink{0009-0006-7540-209X}}
\affiliation{\JKUsemicon}

\author{Thomas Oberleitner\,\orcidlink{0009-0002-0500-2761}}
\affiliation{\JKUsemicon}

\author{Ailton J. {Garcia, Jr.}\,\orcidlink{0000-0002-1364-2622}}
\affiliation{\JKUsemicon}

\author{Saimon Filipe {Covre da Silva}\,\orcidlink{0000-0003-4977-1933}}
\affiliation{\JKUsemicon}\affiliation{\UNICAMP}

\author{Josef Resl\,\orcidlink{0009-0002-6291-2198}}
\affiliation{\JKUlightmatter}

\author{Raphael Joos\,\orcidlink{0000-0001-6147-3926}}
\affiliation{\UniS}

\author{Robert Sittig\,\orcidlink{0000-0002-8912-9008}}
\affiliation{\UniS}

\author{Michael Jetter\,\orcidlink{0000-0002-1311-6550}}
\affiliation{\UniS}

\author{Simone L. Portalupi\,\orcidlink{0000-0003-0012-4073}}
\affiliation{\UniS}

\author{Peter Michler\,\orcidlink{0000-0002-2949-2462}}
\affiliation{\UniS}

\author{Armando Rastelli\,\orcidlink{0000-0002-1343-4962}}
\affiliation{\JKUsemicon}



\input{0_abstract}

\maketitle

\begingroup
\renewcommand{\thefootnote}{*}
\footnotetext{These authors contributed equally to this work.}
\endgroup

\input{1_introduction}
\input{2_design_antenna}
\input{3_diodes}
\input{4_piezo_tuning}
\input{5_improvements}
\input{6_discussion}









\FloatBarrier
\clearpage

\begin{strip}
\hrule
\vspace{0.5em}
\centering
\textbf{Appendix}
\vspace{1em}
\end{strip}

\appendix
\numberwithin{equation}{section}
\numberwithin{figure}{section}
\numberwithin{table}{section}

\input{A_appendix}


\medskip
\textbf{Acknowledgements} \par

This project has received funding from the Austrian Science Fund FWF [10.55776/COE1, 10.55776/PIN4389523, 10.55776/FG5], the EU HE EIC Pathfinder challenges action under grant agreement No. 101115575, and from the QuantERA II program that has received funding from the European Union via the project MEEDGARD (FFG Grant No. 906046).
Additional funding was also provided via the projects EQSOTIC and MEEDGARD. These projects were funded within the QuantERA II Programme that has received funding from the EU’s H2020 research and innovation programme under the GA No 101017733, with funding organization BMFTR (with project numbers 16KIS2060K and 16KIS2057K). 
S.M. acknowledges funding support from the National Quantum Mission, an initiative of the Department of Science  Technology, Government of India, MoE-STARS, IISC (STARS-2/2023-0571), India, and Science and Engineering Research Board (SERB) Core Research Grant (Project No. CRG/2023/007444), India.
\FloatBarrier

\bibliographystyle{model1-num-names}

\bibliography{cas-refs}

\end{document}

%% file: 0_abstract.tex
\begin{abstract}
Tunable photonic architectures that improve the extraction efficiency of light from embedded epitaxial quantum dots across a wide spectral range are key enablers for developing bright sources of single and indistinguishable photons. In this study, we experimentally demonstrate planar multilayer antenna structures consisting of epitaxially grown AlGaAs and InGaAs membranes containing quantum dots sandwiched between metallic Au (or Ag) reflector and director layers together with Al$_2$O$_3$ spacer layers. We show that the linewidths and fine-structure splitting of the neutral exciton emission remain comparable to those measured in the corresponding unprocessed samples, demonstrating that the fabrication process preserves the optical quality of the emitters. In addition to broadband operation, we demonstrate that the planar architecture is compatible with electrical tuning via integrated diode structures, and strain tuning using piezoelectric actuators. In spite of limitations related to optical losses in the ultrathin metallic layers, the demonstrated fabrication simplicity, scalability, and compatibility with tunable quantum emitters establish planar antennas as a promising platform for solid-state quantum photonic devices.


\end{abstract}

%% file: 1_introduction.tex
\section{Introduction}

Epitaxially grown quantum dots (QDs) are already widely used as quantum light sources, providing on-demand narrow-linewidth single-photon emission and generation of highly indistinguishable~\cite{Huber-Loyola2026, Holewa2024} or entangled photons~\cite{Zhou2022}. Potential applications include quantum key distribution (QKD)~\cite{bassobasset2021}, quantum repeaters~\cite{Neuwirth2021, Joos2026}, and photonic quantum computing~\cite{Uppu2021}. \ce{GaAs} QDs embedded in \ce{AlGaAs} emitting in the 780 -- 800 nm spectral range are promising due to their compatibility with rubidium-based quantum memories, low fine-structure splitting stemming from high structural symmetry, and possible integration with \ce{AlGaAs} nanophotonic platforms~\cite{Liu2019}. \ce{InAs} QDs emitting at telecommunication wavelength are very appealing for quantum networks since they offer low-loss transmission through the existing optical fibre infrastructure~\cite{Sittig_Stuttgart_MMB,Holewa2025}. \smallskip

Inefficient collection of QD photoluminescence (PL) emission is a main concern for practical applications. In addition to the total internal reflection at the semiconductor-air interface, the large divergence angle of the emitted light in the far field contributes to this limitation. As a remedy, several photonic structures like photonic crystals~\cite{Lodahl2004}, microlenses~\cite{Gschrey2015}, nanowires~\cite{Claudon2010}, cirular Bragg reflectors (CBRs)~\cite{Liu2019}, distributed Bragg reflectors (DBRs) combined with a solid-immersion lens (SIL)~\cite{Reindl2019}, micropillars combined with DBRs~\cite{Reithmaier2004}, and planar antennas~\cite{Huang2021} have been developed to mitigate this issue.\smallskip

Many applications in quantum communication require (i) tuning of the emission wavelength, e.g., for precise spectral alignment with cavities, other emitters, or atomic transitions useful for quantum memories~\cite{Wolters2017, Thomas2024} and (ii) deterministic control of excitonic charge states for blinking suppression and access spin-selective optical transitions for applications such as quantum repeaters~\cite{Neuwirth2021}. Promising practical approaches for the emission wavelength tuning include integrating the sample with piezoelectric materials (strain-tuning) and embedding the quantum dots in p-i-n diodes (electric field tuning). The latter also enables deterministic charging~\cite{Zhai2020} and blinking suppression~\cite{Schimpf2021_blinking}.\smallskip

In this work, we study Yagi-Uda type antennas~\cite{Checcucci2017}, for which a theoretical and experimental proof-of-concept with QDs have been discussed in Ref.~\cite{Huang2021}. This planar design offers simple fabrication compared to most other photonic structures and eliminates the necessity of deterministic positioning during fabrication, since the cavity properties are not position-dependent. Full compatibility with platforms for strain-tuning~\cite{Trotta2012} to reduce the excitonic fine-structure splitting (FSS) is another feature of this architecture.\smallskip

First, it is demonstrated that the linewidth and FSS of neutral excitons measured on a processed planar antenna are comparable with the ones observed on the same sample before processing. This is important, since processing steps may introduce strain and defects, typically leading to a degradation of those quantities. Furthermore, we provide a proof-of-concept demonstration of QD emission tuning in a diode planar antenna by application of a voltage. In another experiment, we show strain-induced tuning of \ce{InAs/GaAs} QDs emitting in the telecom wavelength regime and embedded in a planar antenna integrated on a piezoelectric crystal. While telecom QDs have already been combined with planar antennas~\cite{Yang2020b}, this is, to the best of our knowledge, the first demonstration of combining planar antennas with strain-tuning. Different from the work of Ref.~\cite{Huang2021}, here the \ce{Ag} top layer is covered with a thin \ce{Au} layer in order to protect \ce{Ag} from oxidation~\cite{Yakubovsky2019,Yang2020a}. Current limitations of our presented approach and possible solutions to overcome them are discussed.

%% file: 2_design_antenna.tex
\begin{figure}
    \centering
    \begin{subfigure}{0.48\textwidth}
        \centering
        \includegraphics[width=70mm]{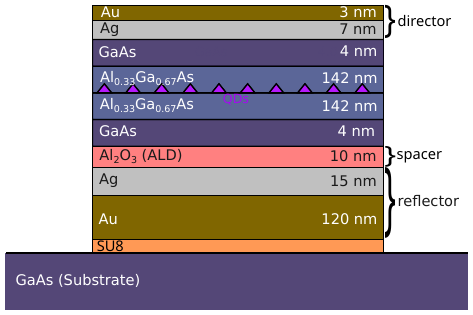}
        \subcaption{}
        \label{fig:F1a}
    \end{subfigure}
    \begin{subfigure}{0.48\textwidth}
        \centering
        \includegraphics[width=80mm]{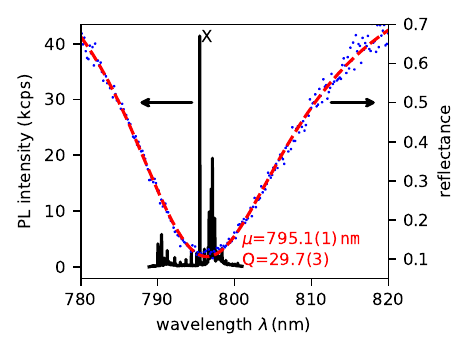}
        \subcaption{}
        \label{fig:F1b}
    \end{subfigure}
    \caption{Planar antenna with GaAs QDs embedded in an intrinsic Al$_{0.33}$Ga$_{0.67}$As membrane. (a) Schematic of the planar antenna design. The \ce{GaAs} quantum dots are indicated by purple triangles. (b) PL spectrum of a representative quantum dot (black) with the neutral exciton (X) spectrally aligned to the cavity mode. The measured normalized reflectance spectrum at \qty{8}{\kelvin} (blue dots) and the corresponding Fano fit (red dashed line) are overlaid.}
    \label{fig:design_choices}
\end{figure}

\section{Planar antenna with narrow-linewidth GaAs QDs in undoped AlGaAs}

The design of the device discussed here is shown in figure \ref{fig:F1a}.  The planar antenna consists of the following stack: a bottom reflector made of \qty{120}{\nano \meter} \ce{Au} and \qty{15}{\nano \meter} \ce{Ag},  \qty{10}{\nano \meter} \ce{Al_2O_3} oxide spacer, \qty{284}{\nano \meter} \ce{Al_{0.33}Ga_{0.67}As} serving as a matrix for the \ce{GaAs} QDs, and a semi-transparent director on top consisting of \qty{7}{\nano \meter} \ce{Ag} and \qty{3}{\nano \meter} \ce{Au}. The \ce{Al_{0.33}Ga_{0.67}As} layers are protected from oxidation on the top and at the bottom by \qty{4}{\nano \meter}-thick \ce{GaAs} capping layers.\smallskip

The GaAs QDs in epitaxially grown \ce{Al_{0.33}Ga_{0.67}As} layer are formed by local droplet etching (LDE)~\cite{Wang2007} and subsequent nanohole filling ~\cite{Heyn2009}. This enables strain-free growth ~\cite{CovreDaSilva2021}. The \ce{Ag} and \ce{Au} layers are formed with physical-vapour deposition (PVD), and the \ce{Al_2O_3} layer is fabricated using atomic layer deposition (ALD). Appendix \ref{sec:appendix_growthstructure} presents the detailed layer sequence of the used wafer, details about the subsequent processing are provided in appendix \ref{sec:appendix_processing}.\smallskip

To better classify the optical stack, we conduct ellipsometry measurements on the oxide film and  metal layers (both for thicker reflector and thinner director) to get the realistic parameters for the dielectric functions of the used materials. The technical details of the ellipsometry study and the simulations are provided in appendices \ref{sec:appendix_optical_investigation} and \ref{sec:appendix_efficiency}, respectively. The collection efficiency depends on the numerical aperture (NA) of the collection lens, values simulated for a given structure are shown in figure \ref{fig:FE1}.\smallskip

The purpose of the \qty{3}{\nano \meter} \ce{Au} layer in the director is to avoid surface oxidation of \ce{Ag}~\cite{Uomoto2016}. Adjusting the thicknesses of the director and the \ce{Al2O3} layer enables tuning the mode position of the optical cavity to the QD emission range between \qty{790}{\nano \meter} and \qty{800}{\nano \meter}. Unlike varying the director layers, which may increase the absorption in those layers, leading to signal losses, adjusting the thickness of the oxide spacer instead allows for positioning the cavity mode without absorption effects. Another important property of the \ce{Al2O3} layer is its ability to passivate the GaAs surface by reducing surface recombination processes~\cite{Frank2005}.\smallskip

The engineered resonance mode around the QD emission at \qty{795}{\nano \meter} leads to a pronounced dip in the reflectance spectrum of the device. Figure \ref{fig:F1b} shows a typical photoluminescence spectrum of a QD collected from an antenna sample together with the relative reflectance dip, observed in the measured white light reflectance spectrum, normalized to the reflection on a \ce{Ag} mirror. The reflectance data is fitted with a Fano resonance model~\cite{Fano1961} tailored for non-symmetrical modes. Technical details of the fitting procedure are outlined in appendix \ref{sec:appendix_optical_investigation}. The mode position is given by \qty{795.1(1)}{\nano \meter}, which matches well with the average wavelength of the QDs emission in this sample. The calculated quality factor (Q factor) of the cavity is $29.7(3)$.\smallskip

\begin{figure}
    \centering
    \begin{subfigure}{0.48\textwidth}
        \centering
        \includegraphics[width=80mm]{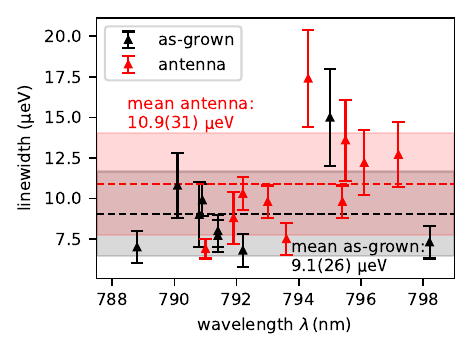}
        \subcaption{}
        \label{fig:F2a}
    \end{subfigure}
    
    \begin{subfigure}{0.48\textwidth}
        \centering
        \includegraphics[width=80mm]{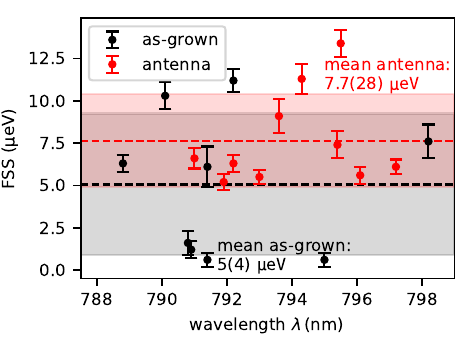}
        \subcaption{}
        \label{fig:F2b}
    \end{subfigure}
    \caption{Comparison of the optical properties of neutral excitons before and after antenna fabrication. (a) Linewidths of representative neutral excitons measured under above-bandgap excitation at \qty{5}{\kelvin} for the as-grown sample (black triangles) and the processed planar antenna (red triangles). (b) Fine-structure splitting (FSS) of representative neutral excitons measured under identical conditions for the as-grown sample (black circles) and the processed planar antenna (red circles). The shaded regions represent the standard deviation of the measured data and do not include the individual measurement uncertainties.}
    \label{fig:comparison}
\end{figure}

The experimental extraction efficiency is determined at a numerical aperture (NA) of the incoupling lens of $0.42$. We employ pulsed two photon excitation (TPE) to coherently prepare the neutral biexciton state.
From the intensity (2.4 kcts/s after losses), together with photon correlation measurements results showing a reduction of efficiency due to blinking by a factor of $0.33(1)$ and a biexciton preparation fidelity of $90\mathtt{\%}$, a value of $3.1(2)\mathtt{\%}$ is calculated following appendix \ref{sec:appendix_efficiency}. A simulation of extraction efficiency, shown in appendix \ref{sec:appendix_efficiency}, leads to an expected efficiency of \qty{5.8}{\percent} at around \qty{795}{\nano\meter} for an NA of $0.42$ and a significant increase for higher values of NA. The mismatch between experiment and simulation is attributed to the difficulty of accurately fabricating thin director layers, particularly our protective \ce{Au} layer. A proposed way to improve the optical performance in future devices  without sacrificing the chemical stability of the director is provided in section \ref{sec:improvements}.\smallskip

Based on time-resolved photoluminescence measurements, the lifetimes under TPE are estimated to be $105\,\mathtt{ps}$ for the biexciton (XX) emission lines and $212\,\mathtt{ps}$ for exciton (X) emission lines. 
This leads to Fourier limits for the linewidth of \qty{9.4}{\mu\eV} for XX and \qty{3.1}{\mu\eV} for the X lines.
We compare this lower limit to the measured photon linwidth of the neutral exciton, obtained via above band gap excitation with a \qty{532}{\nano \meter} laser and  Michelson interferometry for both the as-grown sample before processing and the processed planar antenna. 
The corresponding values as a function of emission wavelength are shown in figure \ref{fig:F2a}. The smallest linewidths found are $6.8(5)\,\mathtt{\mu eV}$ for the unprocessed sample and $6.7(5)\,\mathtt{\mu eV}$ on the processed planar antenna. From the statistics, we deduce a mean linewidth of \qty{9.1(2.6)}{\mu eV} for the unprocessed sample, and  \qty{10.9(3.1)}{\mu eV} for the planar antenna. The standard deviation accounts for the variability between the individual dots, and does not include the uncertainties of individual measurements. Hence, we measured the mean linewidth to be a factor of $3.5(5)$ above the Fourier-transform limit for X, with the lowest linewidth reaching a factor of $2.1(1)$ above the Fourier-transform limit. By comparing our lifetimes with data from other studies, for instance on DBRs~\cite{CovreDaSilva2021}, we can conclude that we see no significant Purcell enhancement, compatible with the simulation results (not shown).\smallskip

The fine-structure splitting (FSS) of neutral exciton lines is analyzed for the same QDs and shown in figure \ref{fig:F2b}. For the planar antenna, a mean FSS of \qty{7.7(2.8)}{\micro \eV} was obtained. We compare that with a mean FSS of \qty{5(4)}{\mu \eV} for the unprocessed sample. The uncertainties again reflect the spread, without the individual uncertainties from the fits. We conclude that the fabrication process does not introduce significant anisotropic stress, which would otherwise tend to increase the FSS more extensively.\smallskip

%% file: 3_diodes.tex
\section{Planar antenna with GaAs QDs in an AlGaAs p-i-n diode}

\begin{figure}
    \centering
    \begin{subfigure}{0.48\textwidth}
        \centering
        \includegraphics[width=70mm]{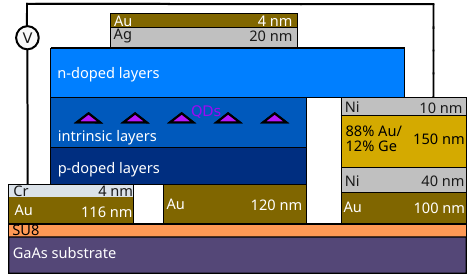}
        \subcaption{}
        \label{fig:F3a}
    \end{subfigure}
    
    \begin{subfigure}{0.48\textwidth}
        \centering
        \includegraphics[width=75mm]{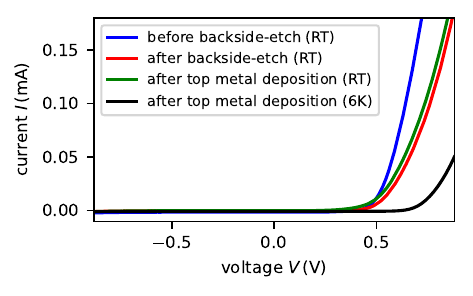}
        \subcaption{}
        \label{fig:F3b}
    \end{subfigure}

    \begin{subfigure}{0.48\textwidth}
        \centering
        \includegraphics[width=80mm]{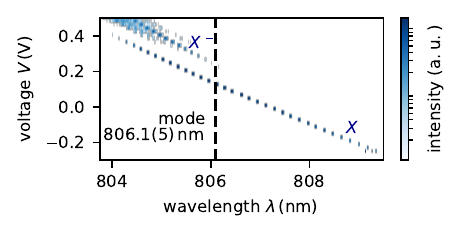}
        \subcaption{}
        \label{fig:F3c}
    \end{subfigure}
    \caption{Planar diode antenna incorporating GaAs quantum dots. (a) Schematic of the fabricated layer structure. (b) Current--voltage characteristics of the diode measured at room temperature (RT) before backside etching, after backside etching, and after deposition of the metallic director layers. The corresponding current--voltage characteristic measured at \qty{6}{\kelvin} is included for comparison. (c) Voltage-dependent PL spectra of a representative quantum dot measured at \qty{6}{\kelvin}, illustrating charge-state control and wavelength tuning via the quantum-confined Stark effect. The cavity-mode position is indicated by the dashed line.}
    \label{fig:design_choices_diode}
\end{figure}

Here we demonstrate the successful fabrication of p-i-n diodes with ohmic contacts integrated in planar antennas. To the best of our knowledge, this approach has not been studied before. Figure \ref{fig:F3a} shows the basic design of the fabricated diode device. The planar antenna consists of \qty{20}{\nano \meter} \ce{Ag} and \qty{4}{\nano \meter} \ce{Au} as director, as well as of \qty{120}{\nano \meter} \ce{Au} for the reflector. The semiconductor region consists of \qty{102}{\nano \meter} n-doped, \qty{123}{\nano \meter} intrinsic, and \qty{64}{\nano \meter} p-doped \ce{AlGaAs} layers. The doped layers also include \qty{4}{\nano \meter} \ce{GaAs} capping layers. Details on the doping levels and \ce{Al} fraction $x$ in \ce{Al_{x}Ga_{1-x}As} layers are provided in appendix \ref{sec:appendix_growthstructure}. The processing work flow for this device is provided in appendix~\ref{sec:appendix_processing}.\smallskip

The diode electrical contacts are formed independently from the metal layers of the optical cavity system. For the p-contact, we use \qty{115}{\nano \meter} \ce{Au} and \qty{5}{\nano \meter} \ce{Cr}, following \cite{Shatalina2010}. The n-contact is formed by the deposition of  \qty{10}{\nano \meter} \ce{Ni}, 150 nm alloy of \qty{88}{\percent} \ce{Au} and \qty{12}{\percent} (\ce{Ge}), \qty{40}{\nano \meter} \ce{Ni}, 120 nm \ce{Au} metal stack and subsequent rapid thermal annealing at \qty{420}{\celsius} for 2 minutes in forming gas atmosphere, which is well-established recipe to form ohmic contacts on GaAs and \ce{GaAs}/\ce{AlGaAs} materials  ~\cite{Bruce1987,Abhilash2010}. The annealing is done before the bonding step, so the contacts remain ohmic on the membrane, unlike former approaches with diode membranes having Schottky contacts~\cite {Aberl2017}.\smallskip

The current-voltage characteristics (IVC) of the diode device were investigated at different stages of fabrication. Figure \ref{fig:F3b} shows the IVC at room temperature (RT) before backside etching (see appendix \ref{sec:appendix_processing}), after backside etching and after deposition of the director layers. The diode characteristics does not change after top metal deposition. Hence, we conclude that the director layers do not interfere with the electrical properties, making the planar antenna design compatible with p-i-n diodes. The IVC at \qty{6}{\kelvin}  taken for up to \qty{50}{\mu A}, is shown in the inset of figure \ref{fig:F3b}, providing a reasonable low-current operating window and proving the stable diode performance at cryogenic temperatures.\smallskip

Voltage-dependent PL measurements demonstrate the wavelength tunability and the QD charge state switching. Figure \ref{fig:F3c} shows the emission of a QD in the diode antenna recorded at \qty{6}{\kelvin} as a function of wavelength for different applied voltage values from \qty{-0.3}{\volt} to \qty{0.5}{\volt}. As expected, the emission lines blue-shift with increasing voltage due to the QCSE. The observed tuning rate of \qty{8.3} nm/V enables to shift the excitonic lines over the cavity mode with FWHM of \qty{14.1(1)}{\nano\meter} and position at \qty{806.1(5)}{\nano\meter}.

%% file: 4_piezo_tuning.tex
\section{Strain-tunable planar antenna with InGaAs QDs}

To demonstrate the compatibility of planar antennas with strain-based wavelength tuning, we investigate a device incorporating telecom-wavelength \ce{InAs/GaAs} QDs integrated on a piezoelectric actuator. Strain tuning provides a versatile approach for spectral alignment of quantum emitters while preserving the possibility of independent electrical control in future device architectures.\smallskip

\begin{figure}
    \centering
    \begin{subfigure}{0.48\textwidth}
        \centering
        \includegraphics[width=80mm]{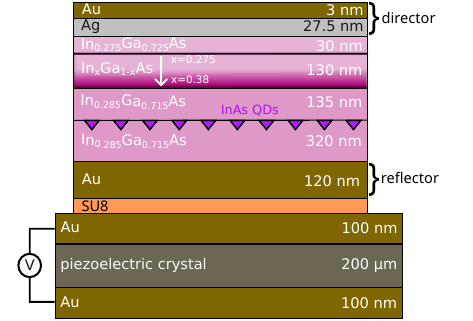}
        \subcaption{}
        \label{fig:F4a}
    \end{subfigure}
    \begin{subfigure}{0.48\textwidth}
        \centering
        \includegraphics[width=80mm]{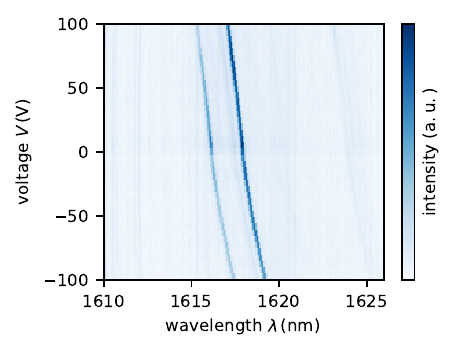}
        \subcaption{}
        \label{fig:F4b}
    \end{subfigure}
    \caption{Strain-tunable planar antenna incorporating \ce{InAs} quantum dots. (a) Schematic of the fabricated layer structure. (b) Colour-coded PL spectra illustrating the strain-induced tuning of the QD excitonic emission under different voltages applied to the piezoelectric actuator.}
\end{figure}

The device structure is shown in figure~\ref{fig:F4a}. The antenna is based on a MOVPE-grown InGaAs metamorphic heterostructure containing InAs quantum dots emitting in the telecom wavelength range. The active region consists of InAs QDs embedded in an \ce{In_{0.285}Ga_{0.715}As} membrane grown on a graded \ce{In_{x}Ga_{1-x}As} metamorphic buffer (MMB) layer, in which the indium content is gradually varied from $x=0.38$ to $x=0.275$. Such graded metamorphic structures are commonly employed to accommodate the lattice mismatch between the substrate and the active region while simultaneously enabling long-wavelength emission from InAs quantum dots through strain engineering and bandgap reduction~\cite{Sittig_Stuttgart_MMB}. The planar antenna is formed by a 120~nm Au reflector and a 27.5~nm Ag / 3~nm Au director layer. The complete structure is bonded onto a \qty{200}{\micro\meter} thick \ce{[Pb(Mg_{0.33}Nb_{0.67})O_{3}]_{0.72}-[PbTiO3]_{0.28}}  (PMN-PT) piezoelectric actuator, enabling the application of controllable biaxial strain. Details of the epitaxial layer sequence and fabrication procedure are provided in Appendices~\ref{sec:appendix_growthstructure} and~\ref{sec:appendix_processing}.\smallskip

The strain-induced tuning behaviour was investigated by recording the PL spectra of a representative QD while varying the voltage applied to the piezoelectric actuator between $-100$~V and $+100$~V. Figure~\ref{fig:F4b} shows the corresponding color-coded spectra obtained under CW above-band excitation. Two dominant emission lines are observed in the wavelength range between 1615~nm and 1620~nm. Over the investigated voltage range, a total wavelength shift of 2.03(7)~nm is achieved, corresponding to a tuning rate of 10.15~pm/V (or equivalently 4.8~$\mu$eV/V).\smallskip

These results demonstrate that the planar antenna architecture is fully compatible with strain-tuning approaches and can be readily extended to quantum emitters operating in the telecom wavelength regime.

%% file: 5_improvements.tex
\section{Discussion and potential routes to performance improvement}\label{sec:improvements}

The present work demonstrates that planar antennas provide a versatile platform for quantum photonic devices by combining efficient broadband light extraction with compatibility for both electrical and strain tuning. Importantly, the fabrication process preserves the optical quality of the embedded quantum dots, as evidenced by the comparable linewidths before and after processing (mean values of $9.1(26)~\mu$eV and $10.9(31)~\mu$eV, respectively) and only a minor increase in the average fine-structure splitting from $5(4)~\mu$eV to $7.7(28)~\mu$eV. Together with the successful realization of electrically contacted p--i--n diodes and strain-tunable telecom quantum dots, these results demonstrate the flexibility of the planar antenna concept for scalable quantum light sources operating across different material platforms and wavelength ranges.

\begin{figure}
    \centering
    \includegraphics[width=70mm]{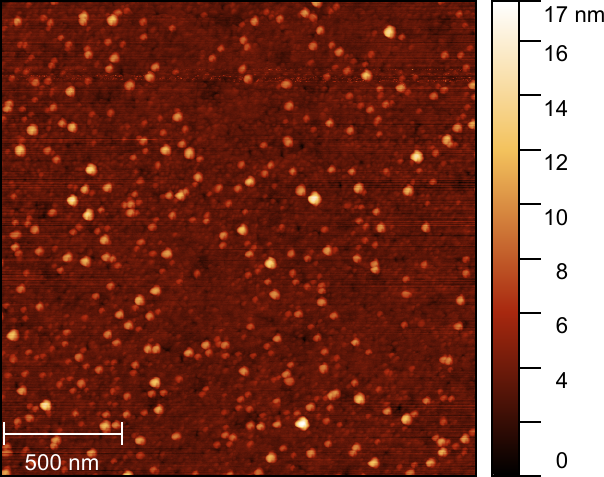}
    \caption{AFM topography of the Ag/Au director layer. Atomic force microscopy image of the sample surface after deposition of a \qty{7}{\nano\meter} \ce{Ag} layer capped with \qty{3}{\nano\meter} \ce{Au} on a \ce{GaAs} substrate, illustrating the morphology of the metallic director layer.}
    \label{fig:F5}
\end{figure}

Despite these promising results, a difference remains between the experimentally measured and simulated extraction efficiencies. For an objective with 0.42 NA, an extraction efficiency of 3.1(2)\% is measured experimentally, compared with a simulated value of 5.8\%. The structural and optical characterization presented in Appendices~\ref{sec:appendix_AFM} and~\ref{sec:appendix_optical_investigation} indicates that this difference originates primarily from the properties of the ultra-thin metallic director rather than from the planar antenna concept itself. AFM measurements reveal RMS surface roughness values of approximately $0.82$~nm for the Ag layer and $1.3$~nm for the Ag/Au director (shown in figure~\ref{fig:F5}), with characteristic grain sizes of about $22$~nm and $19$~nm, respectively. Moreover, the effective Au thickness extracted from the reference samples is approximately $2.5$~nm, suggesting that the nominally 3~nm Au capping layer may not form a fully continuous film. At these dimensions, small variations in morphology and thickness can substantially modify the optical response of the director through increased scattering and absorption. In addition, studies show strong plasmonic effects in a system of thin-film \ce{Au} and \ce{Ag} on \ce{SiO_2}~\cite{Zhong2016, Axelevitch2012} and may also contribute to the additional
losses observed in the present structures.\smallskip

Spectroscopic ellipsometry further confirms that the optical constants of ultra-thin metallic films differ significantly from those of bulk materials. At 795~nm, the imaginary part of the refractive index increases from $k=4.87$ for thick-film Au to $k=7.06$ for the ultra-thin Au layer, indicating considerably stronger optical absorption. Using these experimentally determined optical constants, transfer-matrix simulations predict that approximately $70\%$ of the incident excitation light is absorbed within the Ag/Au director layers. These observations account for the mismatch between simulated and experimentally measured extraction efficiencies and highlight the importance of incorporating experimentally determined optical constants of ultra-thin metallic films into optical simulations of planar antenna structures.

The present study also identifies several realistic routes towards further improving device performance. Replacing the Ag/Au bilayer with Ag--Au alloys containing approximately $95\%$ Ag and $5\%$ Au may preserve the favourable optical properties of Ag while providing enhanced chemical stability ~\cite{Ewald2025} and reduced parasitic absorption. Alternatively, recently developed single-crystalline Ag microflakes ~\cite{Liu2022} offer exceptionally low optical losses and reduced grain-boundary scattering, although at the expense of a more complex fabrication process. Additional improvements are expected from further optimization of the Ag and Au layer thicknesses, deposition conditions to promote smoother and more continuous ultrathin films, and dielectric encapsulation strategies that reduce the required Au capping thickness while maintaining long-term stability. Finally, further optimization of the cavity geometry and dielectric spacer thicknesses offers additional opportunities to enhance the collection efficiency without altering the underlying planar antenna architecture. Altogether, these results indicate that the present performance is limited primarily by the current implementation of the ultra-thin metallic director rather than by the planar antenna concept itself, suggesting that substantially higher extraction efficiencies should be achievable through continued materials and process optimization.

%% file: 6_discussion.tex
\section{Conclusion}

In conclusion, we have demonstrated planar metal-semiconductor dielectric Yagi-Uda type antenna platforms incorporating epitaxial QDs together with electrical and strain-based tuning approaches. The investigated structures support broadband operation and are compatible with scalable fabrication without requiring deterministic positioning of individual emitters.\smallskip

A central result of this work is that the fabrication process preserves the optical quality of the embedded QDs. In particular, linewidths and fine-structure splitting values measured after fabrication remain comparable to those obtained from the corresponding as-grown samples. This demonstrates that the planar antenna approach is well suited for applications requiring narrow-linewidth and low-FSS quantum emitters.\smallskip

Furthermore, we experimentally demonstrated the integration of planar antennas with p-i-n diode structures enabling voltage-controlled tuning via the quantum-confined Stark effect, as well as integration with piezoelectric actuators for strain tuning of telecom-wavelength quantum dots. These results highlight the versatility of the planar antenna platform and its compatibility with different approaches for spectral tuning and charge-state control.\smallskip

Although the experimentally obtained extraction efficiency remains below the ideal simulated values, our investigations identify ultrathin metallic director layers as the dominant source of additional optical losses. In particular, absorption, surface roughness, discontinuous thin-film formation strongly influence the device performance. The combined optical, morphological, and simulation studies presented here provide practical design guidelines for future optimization of planar antenna structures.\smallskip

The presented architecture therefore presents a promising platform for tunable solid-state quantum light sources that combine broadband operation, fabrication simplicity, and compatibility with electrically and strain-tunable quantum emitters. Future optimization of the metallic layers and cavity design may enable significantly improved extraction efficiencies while maintaining the favourable optical properties demonstrated in this work.

%% file: A_appendix.tex
\section{Structure of the as-grown samples}\label{sec:appendix_growthstructure}

Figure \ref{fig:as_grown_structures} summarizes the epitaxial layer structures used for fabrication of the different planar antenna devices investigated in this work.

\subsection{Undoped planar antenna with GaAs quantum dots}

The nominally intrinsic semiconductor antenna is based on an MBE-grown GaAs/AlGaAs heterostructure, shown in figure ~\ref{fig:FA1a}. The structure consists of a \qty{300}{\nano\meter} \ce{Al_{0.75}Ga_{0.25}As} sacrificial layer followed by a \qty{284}{\nano\meter} thick \ce{Al_{0.33}Ga_{0.67}As} membrane containing GaAs QDs. Thin 4~nm GaAs capping layers are included at the top and bottom interfaces of the membrane in order to suppress oxidation of the AlGaAs layers.
The GaAs QDs are fabricated using local droplet etching (LDE) followed by nanohole infilling~\cite{CovreDaSilva2021}, resulting in almost strain-free QDs with high structural symmetry and low fine-structure splitting.

\subsection{Planar diode antenna}

The layer structure used for the planar diode antenna is shown in figure ~\ref{fig:FA1b}. The device is based on a p-i-n \ce{AlGaAs} heterostructure grown by MBE and incorporates \ce{GaAs} QDs embedded within the intrinsic region.\smallskip
The membrane contains n-doped, intrinsic, and p-doped AlGaAs layers with aluminium (\ce{Al}) compositions of \linebreak\ce{Al_{0.15}Ga_{0.85}As} and \ce{Al_{0.33}Ga_{0.67}As}. Silicon is used as the n-type dopant and carbon as the p-type dopant. The nominal doping concentrations are approximately $1 \times 10^{18}$~cm$^{-3}$ for the n-doped layers and $1 \times 10^{19}$~cm$^{-3}$ for the p-doped layers. Thin GaAs capping layers are again included to reduce oxidation effects.

The GaAs QDs are grown using the same local droplet etching and nanohole filling approach as for the intrinsic antenna sample.

\subsection{Strain-tunable planar antenna with telecom quantum dots}

\begin{figure}
    \centering
    \begin{subfigure}{0.425\columnwidth}
        \centering
        \includegraphics[width=32.5mm]{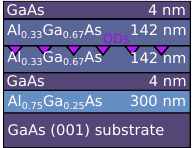}
        \subcaption{}
        \label{fig:FA1a}
    \end{subfigure}
    \hfill
    \begin{subfigure}{0.525\columnwidth}
        \centering
        \includegraphics[width=42.5mm]{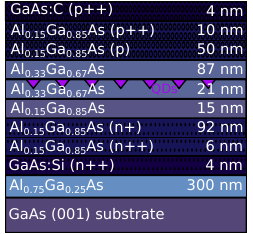}
        \subcaption{}
        \label{fig:FA1b}
    \end{subfigure}
    
    \vspace{0.25cm}
    
    \begin{subfigure}{0.98\columnwidth}
        \centering
        \includegraphics[width=37mm]{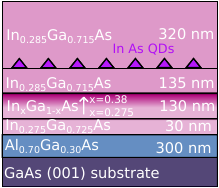}
        \subcaption{}
        \label{fig:FA1c}
    \end{subfigure}
    \caption{Epitaxial layer structures of the as-grown samples. The quantum dots are indicated by purple triangles. (a) Intrinsic semiconductor planar antenna, (b) planar diode antenna, and (c) strain-tunable planar antenna.}
    \label{fig:as_grown_structures}
\end{figure}

The strain-tunable planar antenna is based on a metal-organic vapour phase epitaxy (MOVPE)-grown InGaAs heterostructure containing InAs QDs emitting in the telecom wavelength range, as shown in figure \ref{fig:FA1c}.
The structure consists of a 300~nm Al$_{0.70}$Ga$_{0.30}$As sacrificial layer followed by an InGaAs membrane. The membrane includes a graded In$_x$Ga$_{1-x}$As layer with indium composition varying from $x = 0.275$ to $x = 0.38$, together with a thick In$_{0.285}$Ga$_{0.715}$As region containing the InAs QDs. The vertical position of the QDs is adjusted for maximized coupling to the optical mode of the planar antenna. The heterostructure is designed for subsequent integration with piezoelectric actuators to enable strain-induced wavelength tuning of the QD emission.

\section{Fabrication of planar antennas}\label{sec:appendix_processing}

The fabrication process of the planar antenna devices starts with deposition of the metallic reflector layers and, for the antennas based on undoped layers, the additional Al$_2$O$_3$ spacer layer. The metallic layers are deposited using physical vapour deposition (PVD), while the Al$_2$O$_3$ layer is deposited by atomic layer deposition (ALD). Subsequently, the samples are bonded to 
carrier substrates using SU8 photoresist.\smallskip

After bonding, the GaAs substrate and buffer layers are removed by a sequence of selective wet-chemical etching steps. First, a phosphoric-acid-based solution consisting of 6~ml H$_3$PO$_4$ and 14~ml H$_2$O$_2$ is used for the initial substrate thinning. This is followed by selective etching using a citric acid: H$_2$O$_2$ solution composed of 12~ml of a mixture containing 10.7~g C$_6$H$_8$O$_7$ dissolved in 10.7~g deionized water together with 3~ml H$_2$O$_2$. These etching steps remove the GaAs substrate until the Al$_{0.75}$Ga$_{0.25}$As sacrificial layer is reached. The sacrificial layer is subsequently removed using 1\% HF, resulting in a semiconductor membrane.\smallskip

For the intrinsic antenna and strain-tunable antenna devices, the director layers, consisting of thin Ag and Au films are then deposited by PVD on top of the membrane structure.
Additional processing steps are required for fabrication of the planar diode antenna. To form the n-contact region, the p-doped and intrinsic semiconductor layers are selectively removed using the citric acid: H$_2$O$_2$ etchant. Subsequently, the n-contact metallization consisting of 10~nm Ni/150~nm AuGe (88\% Au, 12\% Ge)/40~nm Ni/120~nm Au is deposited by PVD. Ohmic contact formation is achieved by thermal annealing at 420~$^\circ$C for 2~min in forming gas atmosphere. The p-contact is fabricated by deposition of a 4~nm Cr / 116~nm Au metal stack. 
Finally, the director layer is deposited on the membranes while avoiding overlap between the director and contact regions in order to prevent electrical short-circuiting of the diode structure.

\section{Morphology and thickness investigations with atomic force microscopy}\label{sec:appendix_AFM}

\begin{figure}
    \centering
    \includegraphics[width=80mm]{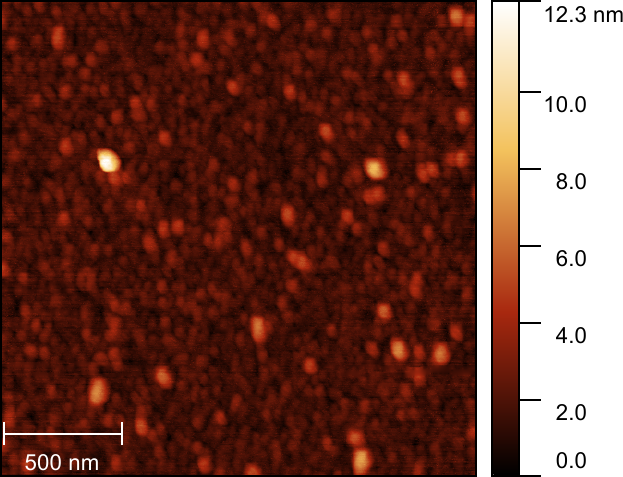}
    \caption{AFM topography of the Ag director layer. Surface morphology of a \qty{7}{\nano\meter} \ce{Ag} film deposited on a \ce{GaAs} substrate, measured by atomic force microscopy.}
    \label{fig:FC1}
\end{figure}

Since the metallic director layers of the planar antennas with undoped materials consist of ultra-thin Ag and Au films, their morphology and thickness strongly influence the optical performance of the device. In order to characterize these layers, dedicated reference samples were fabricated on GaAs substrates by depositing either 7~nm Ag (sample~1) or 7~nm Ag followed by 3~nm Au (sample~2). To enable thickness measurements by AFM, sharp metal edges were defined using photolithography prior to deposition.\smallskip
The fabricated reference samples were used both for morphology investigations and for estimation of the optical parameters employed in the analysis of the spectroscopic ellipsometry (SE) data discussed in Appendix~\ref{sec:appendix_optical_investigation}. Such characterization is necessary because the refractive indices of thin metal films can deviate significantly from their bulk counterparts.\smallskip
Figure~\ref{fig:FC1} and figure~\ref{fig:F5} show representative atomic force microscopy (AFM) images of the Ag and Ag/Au director layers, respectively. From the AFM analysis, root-mean-square (RMS) surface roughness values of \qty{0.82}{\nano\meter} for the \qty{7}{\nano\meter} \ce{Ag} layer and \qty{1.3}{\nano\meter} for the \ce{Ag}/\ce{Au} layer were obtained. Furthermore, granular surface features attributed to island formation are observed, with average grain sizes of \qty{21.8}{\nano\meter} for the \ce{Ag} layer and \qty{18.9}{\nano\meter} for the \ce{Ag}/\ce{Au} layer.\smallskip

Step-height measurements yield layer thicknesses of 5.5(1)~nm for the nominally 7~nm Ag layer (sample~1)and 8.0(5)~nm for the Ag/Au structure (sample~2). Assuming a similar Ag thickness for both samples, the effective Au thickness is estimated to be around 2.5~nm. The comparable magnitude of the Au thickness and surface roughness suggests that the ultra-thin Au layer may not form a fully continuous film. Such non-ideal film formation can increase optical absorption and scattering losses and may additionally reduce the effectiveness of the Au layer in protecting the underlying Ag from degradation.

\section{Investigation of optical properties}\label{sec:appendix_optical_investigation}

Accurate knowledge of the optical properties/refractive indices of all constituent materials is essential for reliable modelling of the planar antenna structures, for instance using finite-difference time-domain (FDTD) simulations or transfer matrix method (TMM) simulation. This includes distinguishing the optical properties of bulk metals with those of thin metals. For \ce{AlGaAs}, we use literature data~ (\cite{Gehrsitz2000} for \ce{Al_{0.33}Ga_{0.67}As}, \cite{Jiang2023} for \ce{GaAs}) and check for consistency. The optical constants of the metallic and dielectric layers were determined by spectroscopic ellipsometry (SE). The extraction efficiency is characterized for the nominally intrinsic semiconductor antenna only, and hence the discussion about optical properties is restricted to the layers for the intrinsic semiconductor antenna. The diode antenna and the strain-tunable antenna are provided as proof-of-concepts. For simplicity, the optical properties are investigated at room temperature only.\smallskip

To determine the optical constants of the reflector materials, spectroscopic ellipsometry (SE) measurements were performed on 150~nm Au, 130~nm Ag, and 160~nm Al$_2$O$_3$ films deposited on Si substrates. To account for the ultra-thin metallic director layers employed in the intrinsic planar antenna, additional SE measurements were carried out on the Ag and Ag/Au reference samples introduced in Appendix~C, consisting of 7~nm Ag on GaAs (sample~1) and 7~nm Ag capped with 3~nm Au on GaAs (sample~2).\smallskip

\begin{figure}
    \centering
    \includegraphics[width=80mm]{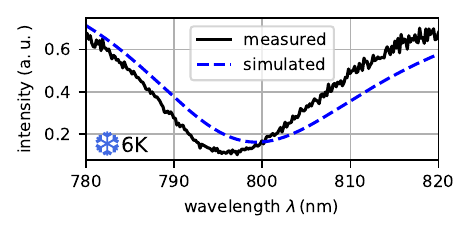}
    \caption{Comparison of measured and simulated reflectance spectra. Measured reflectance spectrum of the intrinsic planar antenna at 6~K together with the corresponding transfer matrix method (TMM) simulation, demonstrating the agreement between experiment and optical modelling.}
    \label{fig:FD1}
\end{figure}

The optical constants of the Ag/Au stack were extracted sequentially by first determining the response of the GaAs substrate, followed by the Ag layer and finally the ultra-thin Au capping layer. Since the Au layer is only 3~nm thick, its optical constants are considerably more sensitive to the fitting procedure than those of the thicker metallic films. Therefore, the SE analysis was further validated by comparing TMM simulations with experimentally measured reflectance spectra of the fabricated antenna structures, as shown in figure \ref{fig:FD1}. In addition, the extracted optical constants were verified to remain physically consistent with the expected increase in both the real ($n$) and imaginary ($k$) parts of the refractive index for ultra-thin Au films compared with bulk material. All SE measurements were performed at room temperature. Nevertheless, the extracted optical constants accurately reproduce the reflectance spectra measured at cryogenic temperature, indicating that temperature-dependent variations are sufficiently small for the present analysis.\smallskip

Within the wavelength range from 775~nm to 870~nm, both the real and imaginary parts of the refractive index exhibit an approximately linear dependence on wavelength. Subsequently, the values of optical constants are tabulated for 795~nm together with the corresponding linear slopes in Table~D.1. The wavelength dependence of the refractive indices is expressed as

\begin{subequations}\label{eq:fit}
    \begin{equation}\label{eq:fit_n}
        n(\lambda) = n(795\mathtt{nm}) + \left(\lambda - 795\right) \cdot n_{\text{slope}},
    \end{equation}
    \begin{equation}\label{eq:fit_k}
        k(\lambda) = k(795\mathtt{nm}) + \left(\lambda - 795\right) \cdot k_{\text{slope}}.
    \end{equation}
\end{subequations}

\begin{table}[ht]
    \centering
    \caption{Optical constants of the investigated materials. The real ($n$) and imaginary ($k$) parts of the refractive index extracted at 795~nm together with the corresponding linear slopes describing their wavelength dependence between 775~nm and 870~nm according to equations (\ref{eq:fit_n}) and (\ref{eq:fit_k}).}
    \label{tab:fit_params}
    
    \begin{subtable}[t]{0.45\textwidth}
    	\centering
    	\caption{thick-film \ce{Au},}
    	\label{tab:Au_bulk_fit_params}
        
        \begin{tabular}{@{}l|c|c@{}}
            \toprule
             $p$ & Value at \qty{795.0}{\nm} & Slope $(dp/d\lambda)$ \\
            \midrule
            n & \qty{0.1209}{\,} & \qty{1.520669e-4}{\per\nm} \\
            k & \qty{4.8700}{\,} & \qty{8.096553e-3}{\per\nm} \\
            \bottomrule
        \end{tabular}
        
    \end{subtable}

    \hfill

    \begin{subtable}[t]{0.45\textwidth}
    	\centering
    	\caption{thin-film \ce{Au},}
    	\label{tab:Au_thin_fit_params}
        \begin{tabular}{@{}l|c|c@{}}
            \toprule
             $p$ & Value at \qty{795.0}{\nm} & Slope  $(dp/d\lambda)$ \\
            \midrule
            n & \qty{0.1391}{\,} & \qty{3.863492e-4}{\per\nm} \\
            k & \qty{7.0612}{\,} & \qty{1.054440e-2}{\per\nm} \\
            \bottomrule
        \end{tabular}
        
    \end{subtable}

    \hfill
    
    \begin{subtable}[t]{0.45\textwidth}
    	\centering
    	\caption{thick-film \ce{Ag},}
    	\label{tab:Ag_bulk_fit_params}
        \begin{tabular}{@{}l|c|c@{}}
            \toprule
             $p$ & Value at \qty{795.0}{\nm} & Slope $(dp/d\lambda)$ \\
            \midrule
            n & \qty{0.0802}{\,} & \qty{9.734418e-5}{\per\nm} \\
            k & \qty{4.6177}{\,} & \qty{6.614901e-3}{\per\nm} \\
            \bottomrule
        \end{tabular}
    \end{subtable}
    
    \hfill
    
    \begin{subtable}[t]{0.45\textwidth}
        \centering
        \caption{thin-film \ce{Ag}, and}
    	\label{tab:Ag_thin_fit_params}    
        \begin{tabular}{@{}l|c|c@{}}   
            \toprule
             $p$ & Value at \qty{795.0}{\nm} & Slope  $(dp/d\lambda)$ \\
            \midrule
            n & \qty{0.5015}{\,} & \qty{5.328962e-4}{\per\nm} \\
            k & \qty{4.8625}{\,} & \qty{6.512180e-3}{\per\nm} \\
            \bottomrule
        \end{tabular}
    \end{subtable}

    \hfill
        
    \begin{subtable}[t]{0.45\textwidth}
    	\centering
    	\caption{\ce{Al_2O_3}.}
    	\label{tab:Al2O3_fit_params}
        \begin{tabular}{@{}l|c|c@{}}
            \toprule
             $p$ & Value at \qty{795.0}{\nm} & Slope  $(dp/d\lambda)$ \\
            \midrule
            n & \qty{1.6431}{\,} & \qty{-1.782831e-5}{\per\nm} \\
            k & \qty{0}{\,} & \qty{0}{\per\nm} \\
            \bottomrule
        \end{tabular}
    \end{subtable}
\end{table}

Comparison of the optical constants in Table~D.1 reveals substantial deviations between thick film and ultra-thin metallic layers. In particular, the imaginary part of the refractive index of Au increases from $k$ = 4.87 for the thick film reflector to $k \sim 7.06$ for the ultra-thin director layer at 795 nm. This increase corresponds to significantly enhanced optical absorption and highlights the importance of using experimentally determined thin film optical constants while modelling planar antenna structures. Such effects contribute directly to the mismatch between idealized simulations and experimentally measured extraction efficiencies.\smallskip

To compare the experimentally measured reflectance spectra with the optical model, TMM simulations were performed using the Python library TMM~\cite{Byrnes2016}. The cavity resonance (see figure \ref{fig:F1b}) was characterized by fitting both the spectra in the vicinity of the resonance using a Fano line shape~\cite{Fano1961}. Specifically, the asymmetric cavity response was described by the \textit{Breit-Wigner model}~\cite{Breit1936}. The fit function is given by

\begin{equation}
    f(x; E_{\text{res}}, \Gamma_{\text{res}}, q) = A \cdot \frac{(q \cdot \Gamma_{\text{res}} / 2 + x - E_{\text{res}})^2}{(\Gamma_{\text{res}} /2)^2 + (x - E_{\text{res}})^2},
\end{equation}

where $E_{\mathrm{res}}$, $\Gamma_{\mathrm{res}}$, and $q$ denote the resonance energy, resonance linewidth, and Fano asymmetry parameter, respectively, and $A$ is a scaling factor. We give the resonant position in wavelength, denoted as $\mu$.\smallskip

The cavity quality factor, which characterizes the spectral selectivity of the resonance, is calculated as

\begin{equation}
    Q = \frac{E_{\text{res}}}{\Gamma_{\text{res}}}.
\end{equation}

The fitting procedure was implemented in Python using the \texttt{scipy} optimization package~\cite{2020SciPy-NMeth} together with the Breit--Wigner model available in the \texttt{lmfit} library~\cite{lmfit}.\smallskip

\begin{figure}
    \centering
    \includegraphics[width=80mm]{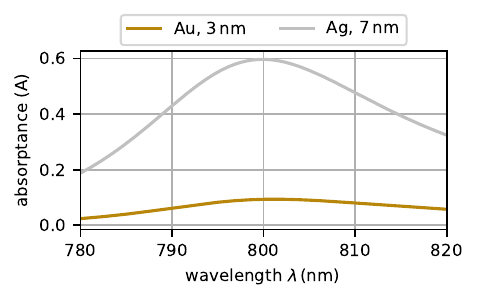}
    \caption{Simulated absorptance of the Ag/Au director layers. Plane-wave transfer matrix method (TMM) simulation of the absorptance $A$ in the \qty{7}{\nano\meter} Ag/\qty{3}{\nano\meter} Au director layers of the intrinsic planar antenna as a function of excitation wavelength, illustrating the parasitic optical losses introduced by the metallic director.}
    \label{fig:FD2}
\end{figure}

To quantify the parasitic optical losses at the excitation wavelength and further assess the role of the metallic director layers in the antenna performance, the absorptance of the individual \ce{Ag} and \ce{Au} director layers was calculated using the transfer matrix method (TMM)~\cite{Harbecke1986} with the same Python library~\cite{Byrnes2016}, assuming coherent plane-wave illumination. The simulated absorptance is presented in figure~\ref{fig:FD2}, revealing that approximately \qty{70}{\percent} of the incident optical power is absorbed within the director layers. This substantial absorption confirms that parasitic losses in the ultra-thin metallic layers constitute one of the dominant factors limiting the extraction efficiency of the planar antenna.

\section{Simulation and measurement of the collection efficiency}\label{sec:appendix_efficiency}

\subsection{Numerical simulation}

The extraction efficiency of the planar antenna structures was estimated numerically using three-dimensional finite-difference time-domain (FDTD) simulations performed with the commercial software package Lumerical FDTD Solutions ~\cite{Othman2022}. The simulations were carried out using the experimentally determined geometrical parameters and optical constants discussed in Appendix~D. The quantum dot emitter was modelled as an electric dipole source positioned at the centre of the semiconductor membrane. Far-field projections were used to evaluate the angular distribution of the emitted radiation and to estimate the fraction of photons collected by an objective lens with a given numerical aperture (NA).\smallskip

The wavelength-dependent extraction efficiency was calculated from the simulated far-field intensity distribution according to ~\cite{Huang2021,Krieger2024},  

\begin{equation}
    \eta_{\lambda} = \frac{\int_{0}^{2\pi} \int_{0}^{\theta_{NA}} |E_{\lambda}(\theta, \phi)|^2 \sin \left( \theta \right) d\theta d\phi}{\int_{0}^{2\pi} \int_{0}^{\pi/2} |E_{\lambda}(\theta, \phi)|^2 \sin \left( \theta \right) d\theta d\phi} \cdot \frac{T_{\lambda}}{F_{P\lambda}}.
\end{equation}

Here, we integrate over the electric far field $E_{\lambda}$ in spherical coordinates over all angles collected by the lens, determined by $\theta_{NA}$, and we divide this by the respective integral over the entire half sphere. This is done for every wavelength $\lambda$ of interest. For symmetry reasons, we start both integrals at $0$. To obtain $\eta_{\lambda}$, we multiply this by the near-field transmittance $T_{\lambda}$ and divide it by the Purcell factor $F_{P\lambda}$.\smallskip

Internally, Lumerical computes the transmitted power using the Poynting vector flux through the monitor surface. The frequency-domain power transmission is computed as ~\cite{lumerical_transmission}:

\begin{equation}
T(f) =
\frac{
\frac{1}{2}
\int_{\mathrm{monitor}}
\mathrm{Re}\!\left(\mathbf{P}(f)\right)\cdot d\mathbf{S}
}
{\mathrm{source~power}(f)} ,
\end{equation}

where $T(f)$ is the normalized transmission as a function of frequency,
$\mathbf{P}(f)$ is the Poynting vector, and $d\mathbf{S}$ represents the
surface normal of the monitor.

The Purcell factor $F_P$ estimates the ratio of the decay rate of the emitter within the planar structure to that of a homogeneous medium. Lumerical internally computes it as ~\cite{lumerical_purcell}: 
\begin{equation}
F_p(f) = \frac{\mathrm{dipole~power}(f)}{\mathrm{source~power}(f)} ,
\end{equation}
where $\mathrm{dipole~power}(f)$ is the actual emitted/radiated power in the simulated environment and $\mathrm{source~power}(f)$ is the amount of power injected into the simulation.\smallskip

\begin{figure}
    \centering
    \includegraphics[width=80mm]{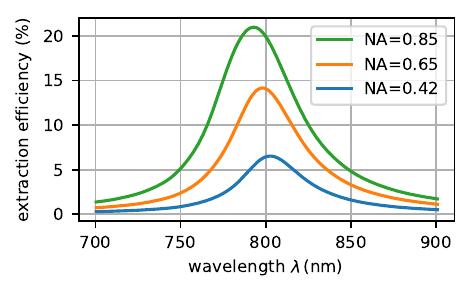}
    \caption{Simulated light collection efficiency. Finite-difference time-domain (FDTD) simulations of the light collection efficiency at 5~K for collection optics with numerical apertures (NAs) of 0.42, 0.65, and 0.85.}
    \label{fig:FE1}
\end{figure}

Figure \ref{fig:FE1} shows a simulation of the extraction efficiency of our intrinsic semiconductor antenna at \qty{5}{\kelvin}. For an NA of $0.42$, we obtain about \qty{5.8}{\percent} extraction efficiency at \qty{795}{\nano\meter}, while for an NA of $0.85$ we expect to achieve a collection efficiency of \qty{20.9}{\percent}.

\subsection{Experimental determination of the extraction efficiency}

The extraction efficiency of the nominally intrinsic planar antenna was experimentally determined using resonant pulsed excitation of a single quantum dot together with a calibrated detection setup. The biexciton state was populated via two-photon excitation (TPE), allowing the number of emitted photons per excitation cycle to be quantified from the measured count rates.\smallskip

The resonant excitation scheme requires efficient suppression of the reflected laser light. For this purpose, a combination of spectral, polarization, and spatial filtering was employed, with the latter implemented using a single-mode optical fibre. Since the fibre coupling efficiency constitutes an additional loss channel, it was determined independently using off-resonant PL measurements. Under excitation with a CW laser at \qty{533}{\nano\meter}, the emitted PL was recorded both in a free-space configuration and after coupling into the single-mode fibre. The integrated intensities were measured using a Teledyne Pylon 100-BRX CCD camera. After correcting for the transmission losses of the respective optical paths, determined using a calibrated laser source and a Thorlabs PM100A power meter, a fibre coupling efficiency of \qty{0.58(5)}{} was obtained.\smallskip

For the resonant excitation measurements, photons were detected using an Excelitas SPCM-AQRH SPAD with a specified detection efficiency of $\eta_{\mathrm{det}}=0.65(2)$. A count rate of 2.4~kcounts/s was measured under pulsed resonant excitation conditions.\smallskip

Several correction factors were taken into account while estimating the extraction efficiency. The influence of blinking was quantified from second-order autocorrelation measurements performed in a Hanbury Brown and Twiss configuration, yielding an on-time of $\beta_{on} = 0.33(1)$. Furthermore, the biexciton preparation efficiency under two-photon excitation was determined from cross-correlation measurements between the biexciton and exciton photons of the radiative cascade, resulting in a preparation fidelity of $\alpha_{prep} = 0.90(2)$.

Combining the measured count rate with the independently determined fiber coupling efficiency, optical transmission losses, blinking probability, and biexciton preparation fidelity, as well as the detector efficiency, yields an extraction efficiency of $3.1(2)\%$ for the investigated intrinsic planar antenna device (figure~\ref{fig:FE1}).